\documentclass[10pt,twocolumn,showpacs,longbibliography,amsmath,amssymb,osajnl,floatfix,superscriptaddress]{revtex4-1}

\usepackage{graphicx}
\usepackage{dcolumn}
\usepackage{bm}
\usepackage{color}
\usepackage{txfonts}
\usepackage{microtype}

\begin{document}
\author{Yan-Fei Li}\email{liyanfei@xjtu.edu.cn}	
\affiliation{Department of Nuclear Science and Technology, Xi'an Jiaotong University, Xi'an 710049, China}
\author{Yue-Yue Chen}\email{yueyuechen@shnu.edu.cn}
\affiliation{Department of Physics, Shanghai Normal University, Shanghai 200234, China}		
\author{Wei-Min Wang}
\affiliation{Department of Physics and Beijing Key Laboratory of
Opto-electronic Functional Materials and Micro-nano Devices, Renmin
University of China, Beijing 100872, China} \affiliation{Beijing National Laboratory for 
Condensed Matter Physics, Institute of Physics, CAS, Beijing 100190,
China}\affiliation{Collaborative Innovation Center of IFSA (CICIFSA), Shanghai Jiao Tong University, Shanghai 200240, China}
\author{Hua-Si Hu}\email{huasi\_hu@mail.xjtu.edu.cn}	
\affiliation{Department of Nuclear Science and Technology, Xi'an Jiaotong University, Xi'an 710049, China}

\title{Production of Highly Polarized Positron Beams via Helicity Transfer from Polarized Electrons in a Strong Laser Field}

\date{\today}

\begin{abstract}

The production of a highly-polarized positron beam via nonlinear Breit-Wheeler processes during the interaction of an ultraintense  circularly polarized laser pulse with a longitudinally spin-polarized  ultrarelativistic electron beam is investigated theoretically. A new Monte Carlo method employing fully spin-resolved quantum probabilities is developed  under the local constant field approximation to include three-dimensional polarizations effects in strong laser fields. The produced positrons are longitudinally polarized through polarization transferred from the polarized electrons by the medium of high-energy photons. The polarization transfer efficiency can approach 100\% for the energetic positrons moving at smaller deflection angles. This method simplifies the post-selection procedure to generate high-quality positrons in further applications. In a feasible scenario, a highly polarized ($40\%-65\%$), intense ($10^5$/bunch$-10^6 $/bunch), collimated ($5$mrad$-70$ mrad) positron beam can be obtained in a femtosecond timescale. The longitudinally polarized positron sources are desirable for applications in high-energy physics and material science .

\end{abstract}

\maketitle
As a powerful probe, spin-polarized positrons play irreplaceable roles in fundamental physical studies and applications. Low-energy (eV to keV) positrons can be utilized to probe the surface \cite{Gidley1982} and  bulk \cite{House1984} magnetism of materials \cite{Rich1987}.  High-energy (GeV to hundreds of GeV) positrons improve the sensitivity of the two photon effect experiments \cite{Elouadrhiri2009},  and are essential for an unambiguous determination of the nucleon structure \cite{Subashiev1998}, testing Standard Model and searching for new physics beyond it \cite{Moortgat2008}. The proposed International Linear Collider (ILC) \cite{Behnke2013} is designed for discovering physics beyond the Standard Model with polarized electrons and positrons at energies of 500 GeV. The positrons are required with polarization more than 30\%, density  $\sim 10^{10}e^+$/bunch, and beam size in nm scale at the interaction point \cite{Flottman1993}.

Polarized positrons can be obtained from beta decays of specific radioisotopes \cite{Zitzewitz1979}. However, the large angular divergence, large energy spread and low intensity of the positron beam from beta decays limit its applications.  
Storage rings can be used to polarized positrons via Sokolov-Ternorv effect \cite{Sokolov1964},  but this time consuming mechanism brings forward rigorous requirements on space scale and layout to experiments. Nowadays, two methods based on BH process are extensively adopted to produce polarized positrons. One is photon-solid interaction, with circularly polarized (CP) $\gamma$ rays generated by linear Compton scattering between CP lasers with unpolarized electrons \cite{Omori2006}, or by synchrotron radiation of unpolarized electrons moving in helical undulators \cite{Alexander2008}. The other is electron-solid interaction, with longitudinally spin-polarized (LSP) electrons \cite{Abbott2016}. However, the energy conversion efficiency from initial electrons to photons in the former way is rather low due to the low fundamental parameter $K (\ll1)$ of the undulator \cite{Corde2013}. 
The latter suffers from high depolarization rates and large angular divergences due to multiple scattering in the Coulomb field of nuclei (Mott scattering) \cite{Potylitsin1997,McMaster1961}, restricting the target thickness to be less than $ 0.2 L_{rad}$ ($ L_{rad}$ is the radiation length typically in several mm
  \cite{Baskov2015}), and consequently limiting the total yield of positrons to $\lesssim 0.01e^+/e^-$ \cite{Potylitsin1997,Abbott2016,Olsen1959}. 
Currently, the state-of-the-art techniques can provide polarized positron beams with polarization $30\% \sim 80\%$,  density $\sim10^{4} e^+/$bunch, and angular divergence  more than 20 degree \cite{Liu2000,Omori2006,Alexander2008,Abbott2016}. Challenging technology upgrades are still needed to meet the experimental requirements above \cite{Flottman1993,Behnke2013}.

Recent progress in development of ultraintense laser system \cite{Yoon2019} has stimulated the interest in producing polarized positrons with strong laser field \cite{Wen2019,Liyf2019,Seipt2019,Wu2019,Song2019,Liyf2020,Chen2019,Wan2020,King2016,Sorbo2017}. 
Since the fierce laser-induced pair production is free from Mott scattering and implemented in nonlinear QED regime ($K\gg1)$), the produced positrons are expected to be a desirable alternative along with outstanding features similar with other laser-driven sources \cite{Esarey2009,Corde2013,Mourou2006,Macchi2013}, such as high brilliance \cite{Ferri2018,Wang2018}, ultrashort duration \cite{Lundh2011}, low angular divergence\cite{Weingartner2012} and high beam intensity \cite{Ma2018,Zhu2016}. 
For instance, an asymmetric two-color laser field can produce polarized positrons with a polarization degree of  around $60\%$, angular divergence of $\sim 74$ mrad and yield of $\sim 0.01 e^+/e^-$ \cite{Chen2019}. Meanwhile, a fine-tuning small ellipticity of a laser pulse results in  an angular dependent polarization of created positrons \cite{Wan2020}. However, all suggested schemes are only able to deliver positrons with transverse polarization, while longitudinal polarization are employed in most applications. To solve this problem, a polarization rotator has to be applied under the risk of particle-amount plummeting since the rotator works for monoenergetic particles with a limited energy range \cite{Steffens1993,Buon1986}. Besides, the effect of photon polarization on pair production is not considered in these shemes.

\begin{figure}[t]
  \setlength{\belowcaptionskip}{-0.1cm}
 	\includegraphics[width=0.95\linewidth]{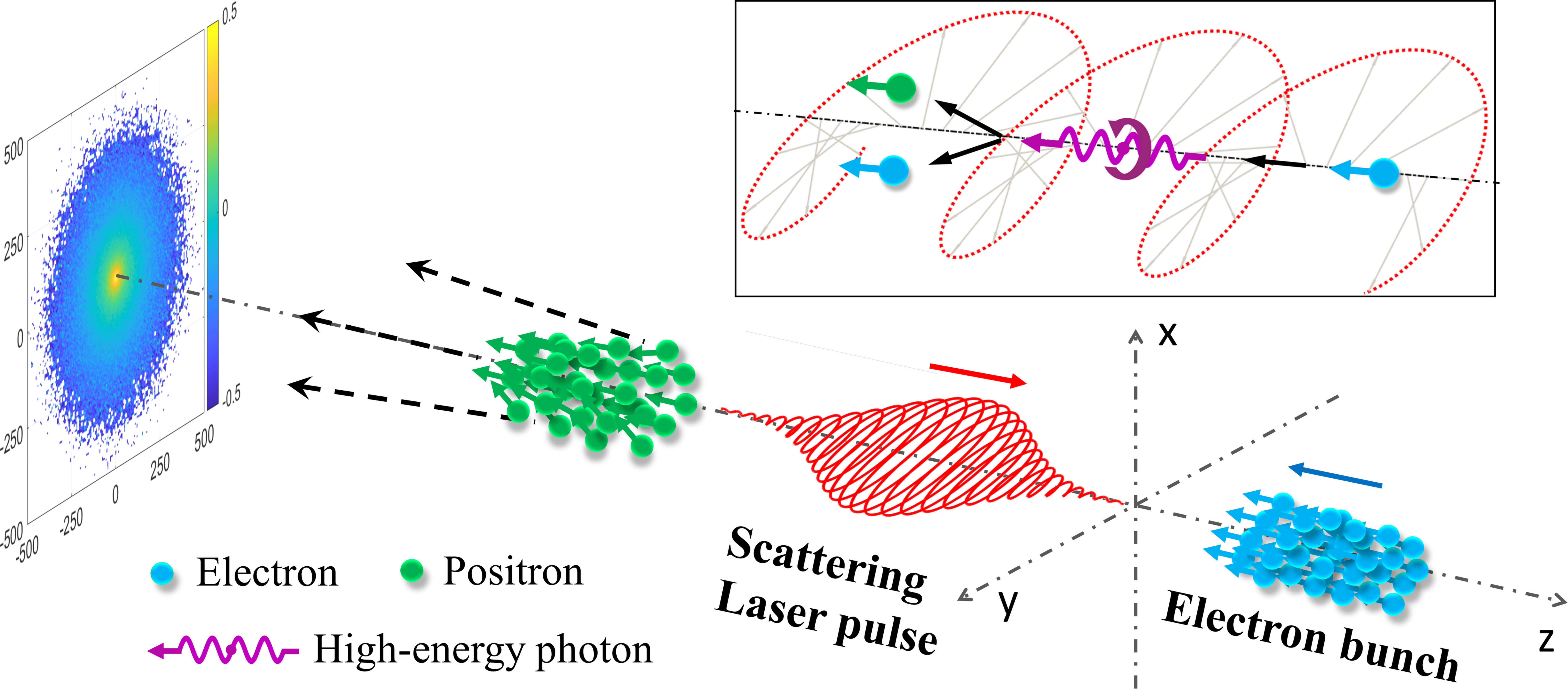}
 \caption{Scenarios of generation of a LSP ultrarelativistic positron beam via an ultraintense laser pulse head-on colliding with a counterpropagating LSP electron beam. First, longitudinal polarization (helicity) is transferred from electron to photon during NCS, then, from high-energy photon to positron through NBW process, as shown in the inset. } \label{fig1}
 \end{figure}

 In this letter, we investigate theoretically the feasibility of production of a longitudinally 
 polarized ultrarelativistic positron beam via the interaction of a CP ultraintense laser pulse with a LSP counterpropagating ultrarelativistic electron beam in the quantum radiation-dominated regime \cite{Piazza2012}, see Fig.~\ref{fig1}. Two steps contribute to the positron polarization. Firstly, circularly polarized photons are radiated during nonlinear Compton scattering (NCS) of a CP laser pulse with a LSP electron beam \cite{Liyf2020}. Then, the helicity of the high-energy photons transfers to electron-positron pairs via nonlinear Breit-Wheeler (NBW) pair production process.  To self-consistently 
incorporate the two processes, a new Monte Carlo for the first time involving all polarization effects from electron (positron) and photon  in realistic tightly-focused laser fields, is developed for simulations. Our simulation shows, under the external electromagnetic fields, a highly polarized intense positron beam can be produced with a small angular divergence. 
 
Our Monte Carlo method \cite{Ritus1985,Plenio1998,Baier1998,Ridgers2014,Elkina2011,Gonoskov2015,CAIN}, treats photon emission and pair production quantum mechanically, and describes the electron (positron) spin-resolved dynamics semiclassically. In particular, photon emission and pair production are conducted by the common statistical event generators, based on quantum probabilities derived via the QED operator method in the local constant field approximation (LCFA) \cite{Baier1998}, to determine whether or not a photon emission or pair production occurs at each simulation step (see the details in the Supplemental Material \cite{SM}). The LCFA is valid in an ultraintense laser fields, with the invariant laser field parameter $a_0 \equiv eE_0/(m\omega_0) \gg 1$, where the formation length of radiation and pair production are far shorter than the laser wavelength and the typical size of the electron (positron) trajectory \cite{Ritus1985,Baier1998,Khokonov2010}. Here, $E_0$ is the laser field amplitude, $\omega_0$ is the laser frequency, and $e (>0),m$ are the electron charge and mass, respectively. Relativistic units $\hbar=c=1$ are used throughout. 
 
Moreover, our Monte Carlo algorithm features the description of polarization effects.
In contrast with the previous Monte Carlo methods related to particular observable of interest \cite{Chen2019,Liyf2019,Liyf2020,Wan2020,Seipt2019,Song2019}, our new method extends the simulation capacity from solving one-dimensional polarization problem to three-dimensional, by choosing the instantaneous spin quantization axis (SQA) according to the properties of the scattering process \cite{Berestetskii1982} instead of  the detector (e.g. the direction of magnetic field in the rest frame \cite{Liyf2019,Chen2019,Wan2020,Seipt2019,Song2019} or the direction of initial electron polarization \cite{Liyf2020}).
The shortcomings of the previous methods stemming from neglecting the phase relation between the two components of the spinor can be overcome \cite{SM}. Meanwhile, since
 photon polarization significantly affects pair production rate ($\geq$ 10\%,  investigated recently in \cite{King2013,Wanf2020}) and positron polarization ($\sim$  60\%, see \cite{SM}), we improved the Monte-Carlo method  
  \cite{Chen2019,CAIN,Wan2020,Elkina2011,Ridgers2014,King2013,Wanf2020} by employing the photon-polarization- and pair-spin-resolved pair production probability applicable to strong laser fields \cite{Baier1998},  and therefore provide a more thorough way to simulate NBW process. 

The details of our Monte Carlo algorithm are elaborated as follow. The electron (positron) spin jumps into one of its basis states defined with respect to SQA in each time step, regardless whether a photon emission happens or not. The spin-resolved radiation probability can be written in form of $W_R={a}+{\bf S}_f^R \cdot {\bf b}$ \cite{Liyf2019,SM}. When a photon emitted,  the  SQA  is chosen to be along ${\bf b}$. The final polarization vector ${\bf S}_{f}^{R}$ is decided with a stochastic procedure: if $W_{R}^+/(W_{R}^++W_{R}^-)>R_a$, ${\bf S}_{f}^{R}=+ {\bf b}/|{\bf b}|$; otherwise ${\bf S}_{f}^{R}=- {\bf b}/|{\bf b}|$. Here, $R_a$ is a random number in [0,1]; and $W_{R}^{+,-}$ are the probabilities calculated by taking ${\bf S}_{f}^{R}$ as $\pm {\bf b}/|{\bf b}|$, respectively. When a photon emission does not occur, the electron (positron) spin should also change quantum mechanically \cite{CAIN}. The probability for no photon emission takes the form of $W_{NR}=\frac{1}{2}(c+\bf{S_f^{NR}}\cdot \bf{d})$ \cite{CAIN}, and the SQA is along ${\bf d}$. The final polarization vector ${\bf S}_{f}^{NR}$ is decided with probability $W_{NR}$ and the stochastic procedure mentioned above. 
 Here, $a$, ${\bf b}$, c and ${\bf d}$ are functions of emitted photon energy, field strength and etc \cite{SM}. The polarization of the emitted photon is determined with a same algorithm \cite{Liyf2020,SM}.

Similarly, the polarization vectors of a newly created pair is calculated with spin-resolved probability of \cite{Baier1998}
\begin{eqnarray}
\label{Wpair}
&&\frac{{\rm d^2}W_{pair}}{{\rm d}\varepsilon_+{\rm d}t}=\frac{C_P}{2}\left(h+{\bf S}_+ \cdot {\bf j}\right),\\
\label{h}
h&=&(\frac{\omega_\gamma^2}{\varepsilon_+\varepsilon_-}-2){\rm K}_{\frac{2}{3}}(\rho)+{\rm IntK}_{\frac{1}{3}}(\rho)-\xi_3{\rm K}_{\frac{2}{3}}(\rho),
\end{eqnarray}
\begin{eqnarray}\label{j}
{\bf j}&=&-\xi_1\frac{\omega_\gamma}{\varepsilon_-}{\rm K}_{\frac{1}{3}}(\rho){\bf \hat e}_1-{\rm K}_{\frac{1}{3}}(\rho)(\frac{\omega_\gamma}{\varepsilon_+}-\xi_3\frac{\omega_\gamma}{\varepsilon_-}){\bf \hat e}_2\nonumber\\
&&+\xi_2\left[\frac{\omega_\gamma}{\varepsilon_+}{\rm IntK}_{\frac{1}{3}}(\rho)+\frac{{\varepsilon_+}^2-{\varepsilon_-}^2}{\varepsilon_+ \varepsilon_-}{\rm K}_{\frac{2}{3}}(\rho)\right]{\bf \hat e}_v
\end{eqnarray}
where $C_p=\alpha m^2/(\sqrt{3}\pi\omega_\gamma)$; $\omega_\gamma$, $\varepsilon_+$ and $\varepsilon_-$ are the energies of the photon, positron and electron, respectively, with $\omega_\gamma=\varepsilon_++\varepsilon_-$; $\rho=2\omega_\gamma^2/(3\chi_\gamma\varepsilon_+\varepsilon_-)$; ${\boldsymbol \xi}=(\xi_1,\xi_2,\xi_3)$ refers to the photon polarization vector with $\xi_{i}$ $(i=1,2,3)$ the Stokes parameters defined with respect to the axes of ${\bf \hat e}_1$ and ${\bf \hat e}_2$ \cite{McMaster1961}; $\hat{{\bf e}}_1$ is the unit vector along the direction of the transverse component of acceleration, $\hat{{\bf e}}_2=\hat{\bf e}_v\times\hat{\bf e}_1$ with $\hat{\bf e}_v$ the unit vector along positron velocity. Quantum parameter is defined as $\chi_{\gamma,e} \equiv |e| \sqrt{-(F_{\mu v}p^v)^2}/m^3$ with $p^v$ the four-vector of photon or electron (positron) momentum.
The spin state of the new-born positron is set as one of the two states: ${\bf S}_+=\pm {\bf j}/|{\bf j}|$, via stochastic procedure. The spin state of produced electron is also obtained with Eq.(\ref{Wpair}), through replacing $\varepsilon_+$,  $\varepsilon_-$ and ${\bf S_+}$ with $\varepsilon_-$, $\varepsilon_+$ and ${\bf S_-}$, respectively.

Between quantum events, the electron (positron) dynamics in the ultraintense laser field are described by Lorenz equations classically, and the spin precession is governed by the Thomas-Bargmann-Michel-Telegdi equation \cite{Bargmann1959}. The detailed description and accuracy of the method are exhibited in the Supplemental Material \cite{SM}. 

  \begin{figure}[htb]
    \setlength{\belowcaptionskip}{-0.1cm}
 	\includegraphics[width=1\linewidth]{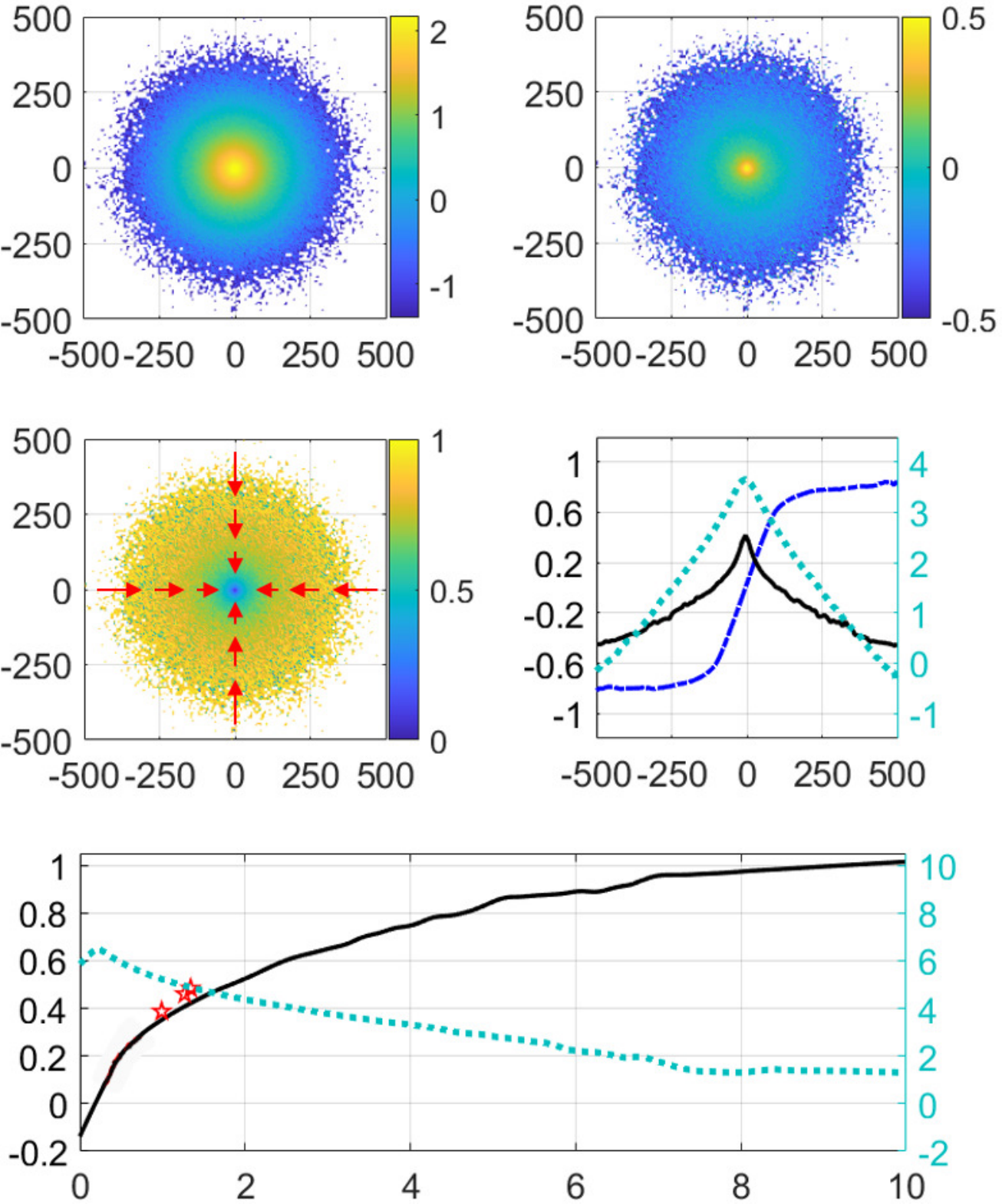}
 \begin{picture}(300,0)
	\put(23,293){(a)}
		\put(56,206){\normalsize $\theta_y$}
	    \put(-5,263){\rotatebox{90}{\normalsize $\theta_x$}} 
	    
	\put(148,293){(b)}
	   \put(180,206){\normalsize $\theta_y$}
	   \put(120,263){\rotatebox{90}{\normalsize $\theta_x$}}	
		
	\put(23,190){(c)}
	    \put(56,103){\normalsize $\theta_y$}
	    \put(-5,160){\rotatebox{90}{\normalsize $\theta_x$}} 
	    
	\put(148,190){(d)}
		\put(180,103){\normalsize $\theta_y$}
	    \put(232,208){\rotatebox{-90}{{\color[rgb]{0,0.75,0.75} log$_{10}$(d$\widetilde N_{e+}$/d$\theta_y$) (mrad$^{-1}$)}}} 
	
	\put(23,86){(e)} 
	    \put(118,150){\rotatebox{90}{$P_\parallel, $ $P_\perp$}} 
		\put(-7,60){\rotatebox{90}{$P_\parallel$}} 
	    \put(235,105){\rotatebox{-90}{{\color[rgb]{0,0.75,0.75} log$_{10}$(d$N_{e+}$/d$\varepsilon_{+}$)  (GeV$^{-1}$)}}} 
		\put(100,2){\normalsize $\varepsilon_{+}$ (GeV)}
	\end{picture}
 \caption{ Angular distribution of number density log$_{10}$(d$^2N_{e+}$/d$\theta_x$d$\theta_y$) (mrad$^{-2}$) (a), longitudinal polarization $P_\parallel=-\overline S_z$ (b), and  transverse polarization degree $|P_\perp|=\sqrt{{\overline S_x}^2+{\overline S_y}^2}$ (c), vs deflection angles of $\theta_x=p_x/p_z$ and $\theta_y=p_y/p_z$. (d) $P_\parallel$ (black-solid), transverse polarization of $P_\perp={\overline S_y}$ (blue-dash-dotted) and positron number density  log$_{10}$(d$\widetilde N_{e+}$/d$\theta_y$) (mrad$^{-1}$) (cyan-dotted) vs  $\theta_y$. Here, d$\widetilde N_{e+}$/d$\theta_y$$=\int_{-20}^{20}{\rm d}^2N_{e+}$/(d$\theta_x$d$\theta_y){\rm d}\theta_x$. (e) $P_\parallel$ (black-solid line) and log$_{10}$(d$N_{e+}$/d$\varepsilon_{+}$) (GeV$^{-1}$) (cyan-dotted line) vs positron-energy $\varepsilon_{+}$. The red stars indicate positrons with deflection angle $\theta=\sqrt{\theta_x^2+\theta_y^2}$ within 5 mrad, 10 mrad, and 20 mrad, from right to left, respetively. } \label{fig2}
\end{figure}
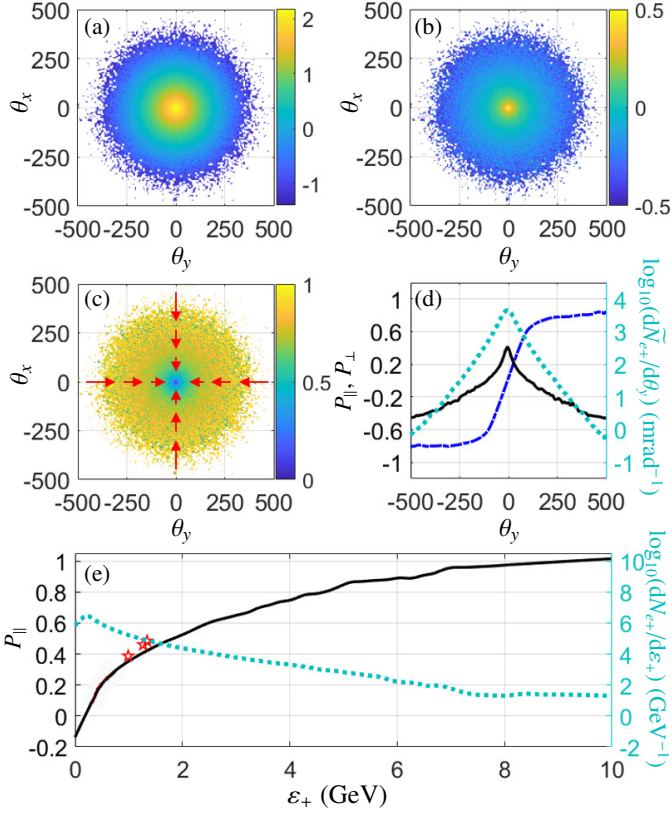
   
  A typical simulation result for production of polarized positrons with a realistic tightly-focused Gaussian laser pulse \cite{Yousef2002} is shown in Fig.\ref{fig2}.  The peak laser intensity is $I_0\approx 2.75\times 10^{22}$ W/cm$^2$ $ (a_0=100\sqrt{2})$, pulse duration (the full width at half maximum, FWHM) 
   $\tau=5T_0$ with $T_0$ the period, wavelength $\lambda=1 \mu$m, and focal radius $w_0=5 \lambda$. The colliding electron bunch is set with features of laser-accelerated electron source \cite{Esarey2009,Gonsalves2019,Leemans2014}. $N_e=9.6\times 10^5$ electrons uniformly distributed longitudinally and normally distributed transversely in a cylindrical form at length of $L_e=6 \lambda$ and
standard deviation of $\sigma_{x,y}=0.6 \lambda$. The initial mean kinetic energy is 10 GeV, the energy spread $6\%$, and the angular divergence 0.2 mrad. The case of a longer electron bunch from traditional accelerators is also considered \cite{SM}.  With the present available electron energy close to 10 GeV by laser wakefield accelerators \cite{Gonsalves2019} and hundreds of GeV by traditional accelerators \cite{Apyan2008}, the laser intensity and electron energy are chosen above to keep $\chi_{\gamma,e}^{max}\approx5.9\times10^{-6}a_0 \gamma_e \gtrsim 1$ for substantial high-energy photon emission and pair production. The initial electrons are set to be 100\% longitudinally polarized, i.e. $S_z=-1$, for a more visible description on polarization transferring  (See \cite{SM} for a more relaxed requirement).
 
The produced positrons mainly concentrate in the center of  the angular distribution with angular divergence (FWHM) around 70 mrad, see Figs.\ref{fig2}(a). The total yield of positrons is 1.17 $e^+/e^-$. Positrons are longitudinally polarized with $P_\parallel>0$ for $\theta\lesssim250$ mrad and $P_\parallel<0$ for $\theta$ $\gtrsim250$ mrad, as shown in Figs.\ref{fig2}(b).  For more intuitive features, one can refer to the angular distribution of density and polarization for positrons  at $\theta_x\in[-20, 20]$ mard, see Fig.\ref{fig2}(d). The positron density dramatically declines with the increase of deflection angle; and $P_\parallel$ decreases with the rising of $|\theta_y|$, from 43\% to  -45\%. 
Moreover, $P_\parallel$ is proportional to positron energy,  similar with that in BH process \cite{Olsen1959}, see Fig.\ref{fig2}(e), but the positron yield is two orders higher. 
Higher polarization can be achieved by using post-selection technique.
For instance, for positrons with energy higher than 2 GeV, 4GeV, 6GeV, and 8GeV, polarization degrees are 62.5\%, 81.9\%, 91.8\% and 98.6\%, respectively, and the corresponding  yields are 0.019 $e^+/e^-$, 0.002 $e^+/e^-$, $1.51\times10^{-4} e^+/e^-$, and $5.21\times10^{-6} e^+/e^-$, respectively. The positron energy ranges from MeV to 10 GeV with a mean value of 0.345 GeV, see Fig.\ref{fig2}(e).
The maximal energy conversion efficiency $\epsilon_{max}\approx1$ and the average energy conversion efficiency $\overline \epsilon\approx 0.034$,  much higher than that in BH process ($\epsilon_{max}\approx0.05$ and $\overline \epsilon \approx 0.003$ for photons from linear Compton scattering \cite{Omori2006}, and $\epsilon_{max}\approx2\times10^{-4}$ and $\overline \epsilon \approx 4\times10^{-5}$ for photons emitted from a electron beam passing through a helical undulator \cite{Dollan2006}).
Besides, the angle-dependent polarization distribution provides a more feasible method to improve polarization by dropping off positrons with higher $\theta$. For instance, the positrons within 5 mrad, 10 mrad, and 20 mrad, have a longitudinal polarization degree of 48.3\%, 46.0\%, and 38.7\%, respectively.
The small emittance ($\sim 0.02$ mm mrad) is
 favorable for experimental operations such as beam injection \cite{Artru2008}.

The positrons have transverse polarization component $P_\perp$ directed radially pointing to the center of the beam-center axis, see Figs.\ref{fig2}(c).  
$P_\perp$ presents an angle dependence as well, i.e.  $P_\perp>0$ for $\theta_y>0$, $P_\perp<0$ for $\theta_y<0$, and the amplitude $|P_\perp|$ increasing with the growing of $|\theta_y|$ from 0 to 80\%, see see Figs.\ref{fig2}(d). 
The radially polarized positron can be used as transversely polarized positrons by collecting positrons in a certain angle. The controlled transverse polarization would be useful for testing detailed structure of the $W^0$ coupling. \cite{Budny1976}.

  \begin{figure}[htb]
 	\includegraphics[width=1\linewidth]{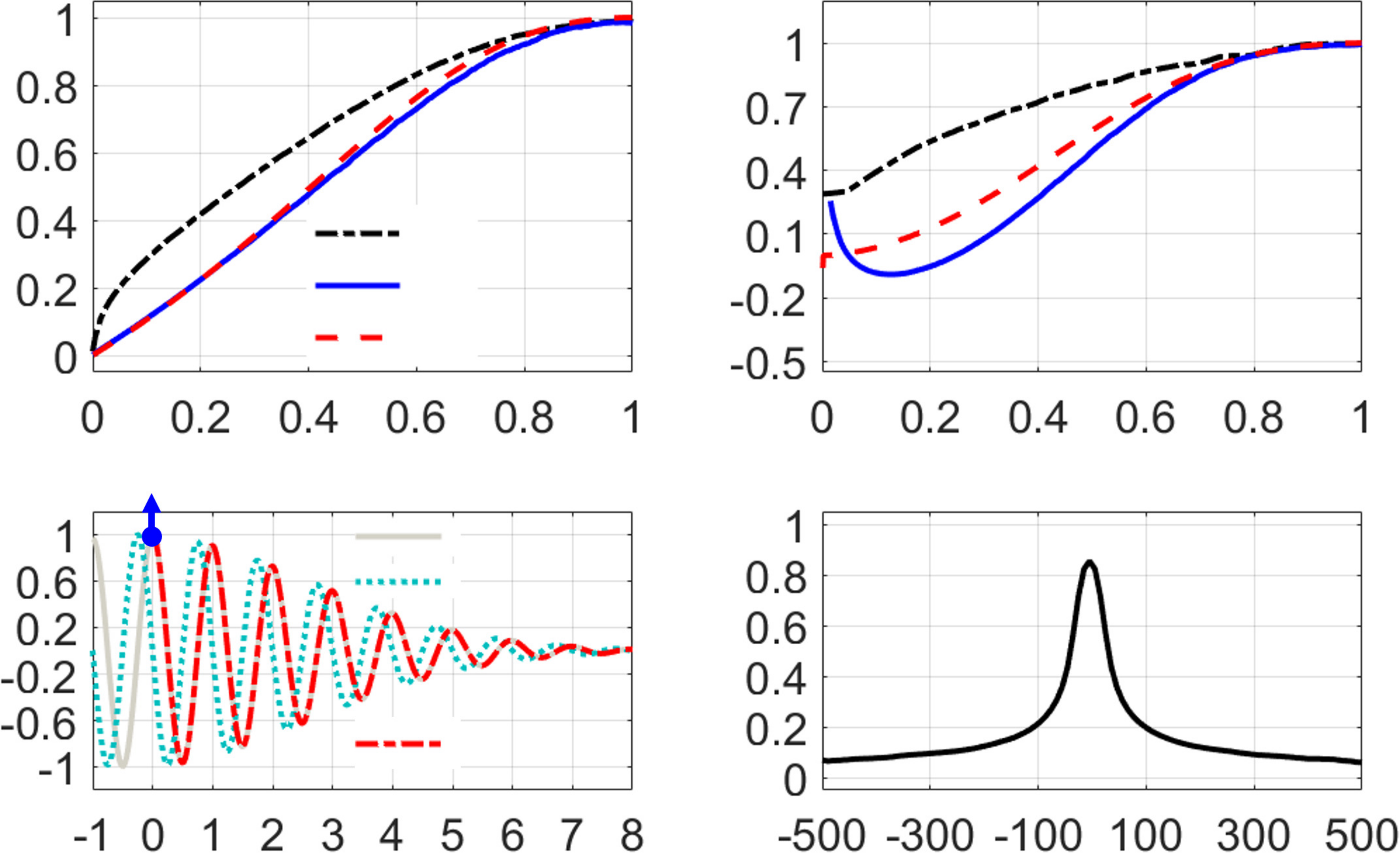}
 \begin{picture}(300,0)
	\put(18,148){(a)}
		\put(61,76){\normalsize $\delta_\gamma$}
	    \put(-15,118){\rotatebox{90}{\normalsize $P_\parallel^{\gamma}$}}
	    \put(71,117){\small with RR }
        \put(71,107.5){\small w/o RR}
        \put(71,98){\small analytical}
	    
	\put(149,148){(b)}	
	    \put(190,76){\normalsize $\delta_+$}
	    \put(117,118){\rotatebox{90}{\normalsize $P_\parallel$}}   

	\put(100,60){(c)}
		\put(41,0){\normalsize $(\eta-\eta_+)/2\pi$}
	    \put(-10,25){\rotatebox{90}{\normalsize abs. units}}
	    \put(78,64){\scriptsize $E_{x'}/E_0$}
        \put(78,55){\scriptsize  $E_{y'}/E_0$}
        \put(78,27){\scriptsize $A_{y'}/A_0$}
	    
	 \put(149,60){(d)}	
	    \put(190,0){\normalsize $\theta_y$}
	    \put(117,28){\rotatebox{90}{\normalsize $\overline \varepsilon_+$ (GeV)}} 
         
	\end{picture}
 \caption{ (a) Circular polarization of photons $P_\parallel^{\gamma}=-\overline \xi_2$ vs  the energy ratio parameter $\delta_\gamma=\omega_\gamma/\varepsilon_i$, from NCS; (b) Longitudinal polarization of positrons $P_\parallel$ vs the energy ratio parameter $\delta_+=\varepsilon_+/\omega_\gamma$,  from NBW of an initial photon beam with $\omega_\gamma=10$ GeV; calculated numerically including (black-dash-dotted) or excluding radiation reaction (RR) effect (blue-solid, using the instantaneous $\chi_{e,\gamma}$ parameter),
 and analytically (red-dashed, employing the constant average value of $ \chi_e=0.97$ or $ \chi_\gamma=4.45$). The other parameters are the same with those in Fig.\ref{fig2}. (c) Normalized field components of $E_{x'}$  (grey-solid), $E_{y'}$ (cyan-dotted) and vector potential $A_{y'}$ (red-dash-dotted), vs laser phase $\eta-\eta_+$, with a positron created at $\eta_+$ (marked with blue point). The blue arrow represents the spin is antiparallel to $\bf{\hat{e}_{y'}}$. (d) Average energy $\overline \varepsilon_+$ vs $\theta_y$, for positrons into $|\theta_x|\le 20$ mrad. (c) and (d) refer to the simulation case in Fig.\ref{fig2}.} \label{fig3}
\end{figure}
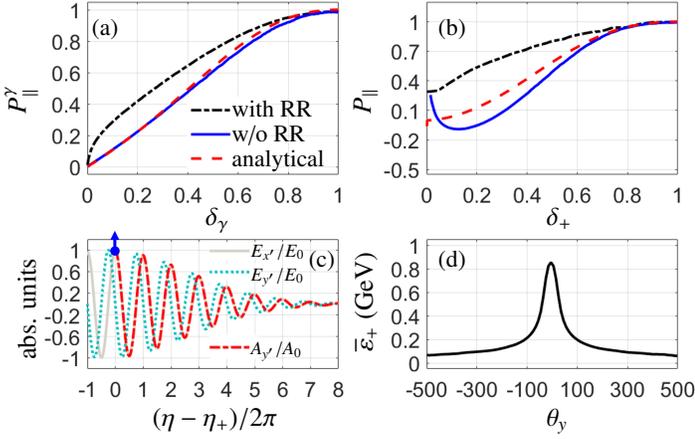

The reason for generating polarized positrons is analyzed in Fig.\ref{fig3}. Processes of photon emission from NCS and pair production from NBW are investigated separately in Figs.\ref{fig3}(a) and \ref{fig3}(b).  
Summing up the final spin states, the analytical estimations on circular polarization of photons and  longitudinal polarization of positrons read,  
\begin{eqnarray}
\label{xi2}
\xi_2 &=& P_\parallel^i \frac{-u{\rm IntK}_{\frac{1}{3}}(u')+u(2+u){\rm K}_{\frac{2}{3}}(u')}{-(1+u){\rm IntK}_{\frac{1}{3}}(u')+2(1+u+u^2/2){\rm K}_{\frac{2}{3}}(u')},\\
\label{sz}
P_\parallel&=& -\xi_2 \frac{\omega_\gamma/\varepsilon_+{\rm IntK}_{\frac{1}{3}}(\rho)+({\varepsilon_+}^2-{\varepsilon_-}^2)/(\varepsilon_+ \varepsilon_-){\rm K}_{\frac{1}{3}}(\rho)}{ (\omega_\gamma^2/(\varepsilon_+\varepsilon_-)-2){\rm K}_{\frac{2}{3}}(\rho)+{\rm IntK}_{\frac{1}{3}}(\rho)-\xi_3{\rm K}_{\frac{2}{3}}(\rho)},
\end{eqnarray} 
where, $P_\parallel^i$ is the initial electron polarization. The numerical results excluding RR effect, in Figs.\ref{fig3}(a) and \ref{fig3}(b), are in coincidence with the analytical ones, with differences mainly coming from the variety of $\chi_e$ ( $\chi_\gamma$) for photons (positrons) created at diverse points in a laser pulse, and (for Fig.\ref{fig3}(b)) from the asymmetry of electromagnetic field experienced by positrons.  When radiation reaction is included, electron (positron) looses energy rapidly. The overlapping of photons emitted by electrons with lower energy ($\varepsilon_t < \varepsilon_i$) at higher $\delta_\gamma^t$ (i.e., higher $P_\parallel^\gamma$), with photons emitted by electrons with $\varepsilon_i$ at $\delta_\gamma^i$, at the condition of $\delta_\gamma^t\varepsilon_t=\delta_\gamma^i\varepsilon_i$, leads to a higher  numerical polarization in the low energy part of $\delta_\gamma\ll1$, in Fig.\ref{fig3}(a). Similarly, the overlapping of positrons created at higher $\delta_+^t$ (i.e., higher $P_\parallel$) experienced more energy-loss $\Delta \varepsilon_+^t$, with positrons created at lower $\delta_+^i$ experienced less energy-loss $\Delta \varepsilon_+^i$, at the condition of $\delta_+^t\omega_\gamma-\Delta \varepsilon_+^t=\delta_+^i\omega_\gamma-\Delta \varepsilon_+^i$, results in a higher  numerical polarization in the low energy part of $\delta_+\ll1$, in Fig.\ref{fig3}(b).
Above all, with helicity transferred from initial electron to photon, then to pair, we acquire the longitudinally polarized positrons in Fig.\ref{fig2}.
The circular polarization of photon (positron) is proportional to its energy, and could approach 100\% as $\omega_\gamma$ ($\varepsilon_+$) gets close to $\varepsilon_i$ ($\omega_\gamma$). Intuitively, the simultaneous energy and helicity transferring, from parent particle to new-born particle, causes the fact that higher helicity transfer efficiency would be accompanied by higher energy transfer efficiency. 

When a $e^+e^-$ pair is created at  laser phase $\eta_+$, the final transverse momentum of the positron is ${\bf p}_\perp^f\approx{\bf p}_\perp^i-e{\bf A}(\eta_+)$, where ${\bf p}_\perp^i$ is the momentum inherited from the parent photon, and ${\bf A}(\eta_+)$ is the vector potential at production point. Since ${\bf p}_\perp^i$ is arbitrary due to the stochastic effects, $\overline{{\bf p}_\perp^\gamma} \approx 0$,  and consequently the final transverse momenta of positrons should be ${\bf p}_\perp^f\approx- e{\bf A}(\eta_+)$. 
For simplicity, we rotate the laboratory coordinate system with respect to instantaneous electromagnetic field, such that $E_{x'}=E_0$, $E_{y'}=0$, as shown in Fig. \ref{fig3}(c).
In the new coordinate system, the final transverse momentum of a positron is antiparallel to the instantaneous ${\bf \hat e}_{y'}$. Since the pairs are mainly created by energetic photons with longitudinal polarization \cite{SM},
the transverse polarization arises from the second term in Eq. (\ref{j}), i.e. ${\rm K}_{\frac{1}{3}}(\rho)\omega/\varepsilon_+{\bf \hat e}_{y'}$, which also indicates polarization degree $|P_\perp|$ inversely proportional to energy. Therefore, positrons are produced with ${\bf p}^f_\perp$ antiparallel and ${P}_\perp$ parallel to $ {\bf \hat e}_{y'}$ with the rotating of $ {\bf \hat e}_{y'}$, i.e. the positrons are polarized radially, as shown in Fig. \ref{fig3}(c). Meanwhile, as deflection angle $\theta=p_\perp/p_\parallel\sim1/\gamma_{e+}$, positrons with higher energy move at smaller deflection angles,
as shown in Fig.\ref{fig3}(d). Above all,  positrons moving closer to axis own higher energy, larger $P_\parallel$ but smaller $P_\perp$, as shown in Fig.\ref{fig2}(d).

The impacts of laser and electron beam parameters on the production of polarized positrons are investigated in \cite{SM}. The polarization of the produced LSP positrons is robust against the variation of pulse duration $\tau (3-8 T_0)$ and peak intensity $a_0 (50\sqrt{2}-100\sqrt{2})$ of the laser pulse, and initial average kinetic energy $\varepsilon_i (5-10 {\rm GeV})$, angular divergence (0.2-5 mrad) and  energy spread (0.06-0.2) of the electron beam.  The scheme works very well even for a long electron bunch ( $L_e=100\lambda_0$).  Generally speaking, larger $a_0$, $\tau$ and $\varepsilon_i$ are conducive to higher yield and energy of photons emitted and positrons created, as $N_{e+}\propto N_\gamma \sim \alpha a_0\tau/T_0$ and $\varepsilon_+\sim \omega_\gamma\sim \chi_e \varepsilon_i \sim 10^{-6}a_0\varepsilon_i^2/m$ \cite{Piazza2012, Ritus1985}. However, since large $\chi_e$ causes strong radiation loss and depolarization \cite{Baier1998}, a trade off exists for $a_0$,$\tau$ and $\varepsilon_i$.

In conclusion, we have proposed a novel method on production of a highly polarized intense ultrarelativistic positron beam via PW laser pulses available recently, with the help of a newly developed Monte Carlo method.  
In a feasible scheme with a seed electron beam with polarization degree 80\%, density  $10^8$/bunch and  kinetic energy 10 GeV \cite{Wen2019,Gonsalves2019}, a high-quality positron beam can be generated with polarization degree 40\%, angle range 5 mrad, density $10^6$/bunch and average energy 1.4 GeV. Given a possible ultrahigh-charge ($\sim100$ nC \cite{Ma2018}) of electron beams, a positron beam density of $10^9\sim10^{10}$/bunch is foreseeable. 
The yield and angular divergence of positron beam is increased and decreased, respectively, by orders with respected to the current available ones, making it a promising alternative source for future experimental facilities in high-energy physics, such as ILC. 
The unavoidable wide energy spread ($\sim 300 $ MeV) could be remedied by post-acceleration, e.g. to less than 0.1\% at 500 GeV. Moreover, the positron beam has a high flux up to $\sim 10^{19}e^+$/s thanks to the ultrashort duration ($L_e\simeq 20$fs), which is favorable for probe \cite{Djourelov2016}  along with a potential for ultrafast diagnosis.\\

{\it Acknowledgement:} The authors thank Y.-T. Li and K. Z. Hatsagortsyan for helpful discussion. This work is supported by the National Natural Science Foundation of China (Grants Nos. 11804269 and Nos. 11775302), the National Key R\&D Program of China (Grant No. 2018YFA0404801), and the Strategic Priority Research Program of Chinese Academy of Sciences (Grant No. XDA01020304).

\bibliography{pair.bib}

\end{document}